\documentclass[usenatbib]{mn2e}
\usepackage{graphicx,natbib}
\usepackage{bm,url,color}
\graphicspath{{./fig/}{./png/}}

\newcommand{\EQ}{\begin{equation}}
\newcommand{\EN}{\end{equation}}
\newcommand{\EQA}{\begin{eqnarray}}
\newcommand{\ENA}{\end{eqnarray}}

\newcommand{\Eqs}[2]{Equations~(\ref{#1}) and~(\ref{#2})}

\newcommand{\Fig}[1]{Fig.~\ref{#1}}

\newcommand{\Figs}[2]{Figs.~\ref{#1} and \ref{#2}}

\newcommand{\Tab}[1]{Table~\ref{#1}}

\newcommand{\bra}[1]{\langle #1\rangle}

{}
{}
{}

{}
{}
{}
{}
{}
{}
{}
{}
{}
{}
{}
{}
{}
{}
{}
{}
{}

{}

{}

{}
{}
{}

\newcommand{\kk}{\bm{k}}

\newcommand{\xx}{\bm{x}}

\newcommand{\rr}{\bm{r}}

\newcommand{\BB}{\bm{B}}

\newcommand{\JJ}{\bm{J}}

\newcommand{\AAA}{\bm{A}}

\newcommand{\uu}{\bm{u}}

\newcommand{\ff}{\mbox{\boldmath $f$} {}}

\newcommand{\nab}{{\bm{\nabla}}}

\newcommand{\SSSS}{\mbox{\boldmath ${\sf S}$} {}}

\newcommand{\ii}{{\rm i}}

\newcommand{\DD}{{\rm D} {}}

\newcommand{\dd}{{\rm d} {}}

\newcommand{\const}{{\rm const}  {}}

\def\Pm{\mbox{\rm Pr}_{\rm M}}
\def\Rm{\mbox{\rm Re}_{\rm M}}

\def\Rmc{\mbox{\rm Re}_{\rm M}^{\rm crit}}
\def\Rey{\mbox{\rm Re}}

\def\cs{c_{\rm s}}

\def\kf{k_{\rm f}}

\def\epsK{\epsilon_{\rm K}}

\def\Brms{B_{\rm rms}}
\def\Bref{B_{\rm ref}}

\def\urms{u_{\rm rms}}

\newcommand{\yapj}[3]{ #1, {ApJ,} {#2}, #3}

\newcommand{\yapjl}[3]{ #1, {ApJ,} {#2}, #3}

\newcommand{\yan}[3]{ #1, {Astron.\ Nachr.,} {#2}, #3}

\newcommand{\yana}[3]{ #1, {A\&A,} {#2}, #3}

\newcommand{\yjfm}[3]{ #1, {J.\ Fluid Mech.,} {#2}, #3}

\newcommand{\ypf}[3]{ #1, {Phys.\ Fluids,} {#2}, #3}

\newcommand{\ypp}[3]{ #1, {Phys.\ Plasmas,} {#2}, #3}

\newcommand{\yjetp}[3]{ #1, {Sov.\ Phys.\ JETP,} {#2}, #3}

\newcommand{\yprl}[3]{ #1, {Phys.\ Rev.\ Lett.,} {#2}, #3}

\newcommand{\ymn}[3]{ #1, {MNRAS,} {#2}, #3}

\newcommand{\ypre}[3]{ #1, {Phys.\ Rev.\ E,} {#2}, #3}

\newcommand{\yjour}[4]{ #1, {#2}, {#3}, #4}

\newcommand{\ybook}[3]{ #1, {#2} (#3)}

\title[Varying the forcing scale in low Prandtl number dynamos]
{Varying the forcing scale in low Prandtl number dynamos}
\author[A. Brandenburg et al.]{
A. Brandenburg$^{1,2,3,4}$\thanks{E-mail:brandenb@nordita.org},
N. E.\ L.\ Haugen$^{5,6}$, Xiang-Yu Li$^{1,2,7}$ and K. Subramanian$^8$
\\
$^1$Laboratory for Atmospheric and Space Physics,
University of Colorado, Boulder, CO 80303, USA\\
$^2$Nordita, KTH Royal Institute of Technology and Stockholm University,
10691 Stockholm, Sweden\\
$^3$JILA and Department of Astrophysical and Planetary Sciences,
University of Colorado, Boulder, CO 80303, USA\\
$^4$Department of Astronomy, Stockholm University, SE-10691 Stockholm, Sweden\\
$^5$SINTEF Energy Research, 7465 Trondheim, Norway\\
$^6$Department of Energy and Process Engineering, NTNU,
7491 Trondheim, Norway\\
$^7$Department of Meteorology and Bolin Centre for Climate Research, Stockholm University, Stockholm, Sweden\\
$^8$Inter University Centre for Astronomy and Astrophysics,
Post Bag 4, Pune University Campus, Ganeshkhind, Pune 411 007, India\\
}

\date{\today,~ $ $Revision: 1.87 $ $}
\begin{document}
\maketitle

\begin{abstract}
Small-scale dynamos are expected to operate in all astrophysical
fluids that are turbulent and electrically conducting, for example the
interstellar medium, stellar interiors, and accretion disks, where
they may also be affected by or competing with large-scale dynamos.
However, the possibility of small-scale dynamos being excited at small and
intermediate ratios of viscosity to magnetic diffusivity (the magnetic
Prandtl number) has been debated, and the possibility of them depending
on the large-scale forcing wavenumber has been raised.
Here we show, using four values of the forcing wavenumber,
that the small-scale dynamo does not depend
on the scale-separation between the size of the simulation domain and the
integral scale of the turbulence, i.e., the forcing scale.
Moreover, the spectral bottleneck in turbulence, which has been implied
as being responsible for raising the excitation conditions of small-scale
dynamos, is found to be invariant under changing the forcing wavenumber.
However, when forcing at the lowest few wavenumbers, the effective
forcing wavenumber that enters in the definition of the magnetic Reynolds
number is found to be about twice the minimum wavenumber of the domain.
Our work is relevant to future studies of small-scale dynamos, of which
several applications are being discussed.
\end{abstract}

\begin{keywords}
dynamo --- MHD -- magnetic fields --- turbulence --- Sun:dynamo
\end{keywords}

\section{Introduction}

Magnetic fields are ubiquitous in astrophysics.
In fact, most of the gas in the universe is ionized and therefore
electrically conducting.
This allows part of the kinetic energy of the gas to be converted into
magnetic energy through the dynamo instability.
This is when the induction equation, which is linear in the
magnetic field $\BB$, has exponentially growing solutions,
starting from just a weak seed magnetic field.
A linear evolution equation can also be formulated for the
two-point correlation function, $\bra{B_i(\xx)B_j(\xx+\rr)}$, where angle
brackets denote averaging, for example over volume, $\xx$ is the position
vector, and $\rr$ the separation between the two points.
Depending on the statistics of the velocity field and the electric
conductivity, this equation may have exponentially growing solutions.
In that case, we talk about a small-scale dynamo instability,
where $\bra{\BB^2}$ grows exponentially, in contrast
to a large-scale dynamo instability where $\bra{\BB}$ itself grows exponentially.

The first rigorous derivation of exponentially growing solutions
for $\bra{\BB^2}$ in statistically mirror-symmetric homogeneous
turbulence goes back to the early work of
\cite{Kaz68}, but it was not until direct numerical simulations
since \cite{MFP81} started seeing small-scale dynamo action on
the computer that this work became more widely known.
Subsequent work, notably by \cite{KA92}, \cite{Sub99}, and \cite{BCR05},
have significantly contributed to our understanding of small-scale
dynamos and their interaction with large-scale ones; see the review
by \cite{BSS12}.
In particular, the presence of flows which allow for a large-scale dynamo
may also ease the excitation of the small-scale dynamo.
This, however, always requires some
departure from the most generic form of turbulence, which is isotropic,
homogeneous, and statistically mirror-symmetric.
Thus, a small-scale dynamo is a generic feature of any turbulence
in electrically conducting gases or fluids, because it is
able to produce dynamically important magnetic energy densities
comparable to the kinetic energy density.
This is when the velocity begins to depend on $\BB$ and the induction
equation is no longer linear in $\BB$, leading to dynamo saturation.
Such a state can be regarded as a generalization of standard
Kolmogorov turbulence to an ionized gas or fluid that is electrically
conducting and therefore subject to the small-scale dynamo instability.
However, how generic this generalization of standard Kolmogorov turbulence
really is depends on how generic is this dynamo instability.

Already over 20 years ago, analytic work showed that the dynamo threshold
of the small-scale 
\cite[also known as fluctuation dynamo; see][]{Scheko07,Eyink10,BS13}
increases with decreasing
magnetic Prandtl number \citep{RK97}, i.e., the ratio $\Pm=\nu/\eta$
of fluid viscosity $\nu$ to magnetic diffusivity $\eta$.
In the interstellar medium, where the density is very low, we have $\Pm\gg1$,
but in stars, and even in their outer layers, we have
$\Pm\ll1$.
The increase of the dynamo threshold with decreasing $\Pm<1$ was then
confirmed numerically by \cite{Scheko04,Scheko05} and \cite{HBD04}.
By solving the \citet{Kaz68} equation,
\cite{Schober} confirmed the increased critical value of $\Rm$ both for
Kolmogorov and Burgers turbulence.
From a technical point of view, the problem is doubly difficult.
On the one hand, as we decrease $\Pm$, even when keeping the magnetic
Reynolds number and therefore $\eta$ unchanged, we already need more
numerical resolution because $\nu$ decreases and therefore the fluid Reynolds
number increases.
On the other hand, if the dynamo threshold is increasing, 
one must increase the magnetic Reynolds number for the dynamo
to remain supercritical, demanding even higher
fluid Reynolds numbers and therefore even higher numerical resolution.

The increase of the dynamo threshold was interpreted as being a
consequence of the scale where the magnetic spectrum peaks
during the kinematic stage, moving away from the
viscous subrange into the inertial range of the turbulence where the
velocity field becomes rougher \citep{BC04}.
Here, rougher means that velocity differences $v_l$ between two points
separated by $l$, scale as $v_l \propto l^{\alpha}$ with $\alpha <1$.
However, we have known for some time that, just before 
exiting
the inertial range, the kinetic energy spectrum becomes even shallower.
This is generally known as the spectral bottleneck effect \citep{SJ93}.
It is theoretically explained as a consequence of a reduced efficiency
of triad interactions with modes in the viscous subrange \citep{Fal94}.
This has consequences for the dynamo threshold.
It makes the velocity even rougher than for Kolmogorov turbulence
(which has $\alpha=1/3$)
near the bottleneck region of the spectrum.
Subsequent work of \cite{Isk07} showed that this particular problem
of the small-scale dynamo suffering from the spectral bottleneck can
be overcome by decreasing the magnetic Prandtl number even further.
The most difficult case is $\Pm\approx0.1$, where the excitation conditions
were so high that nobody was able to confirm that the small-scale dynamo
could be excited.
\cite{Isk07} found supercritical solutions near $\Pm\approx0.1$
when using hyperviscosity in their simulations, which does however
affect the strength of the spectral bottleneck, as will be discussed
in a moment.
Even the recent convection simulations of \cite{KKB18} at a resolution
of $1024^3$ meshpoints found only decaying magnetic fields at $\Pm=0.1$.

In the meantime, two further developments have occurred.
On the one hand, in the nonlinear regime the case $\Pm=0.1$ turned out
to be not so difficult as in the kinematic dynamo regime and sustained
dynamo action has been found.
This could be explained by the spectral bottleneck being suppressed by
a dynamically important magnetic field; see Fig.~2 of \cite{Bra11}
for $\Pm=0.05$ and $0.02$.
For $\Pm=1$, on the other hand, the bottleneck is not yet strongly
suppressed; see Fig.~2 of \cite{HB06}.
On the other hand, even in the kinematic (linear) problem,
\citet*[][hereafter SB14]{SB14} found small-scale dynamo action at
$\Pm=0.1$ in their simulations where turbulence was being forced at a
wavenumber of about four, instead of one to two, which had been used
in all the small-scale dynamo studies until then.
This raised the new possibility that maybe the spectral bottleneck
itself could be an artifact of having forced the turbulence at or near
the scale of the domain.
Clarifying this question is the main goal of the present paper.

An artificially enhanced bottleneck effect has been seen in simulations
that use hyperviscosity instead of the regular diffusion operator
proportional to $\nabla^2$.
\cite{BM00} found that this artificially produced bottleneck effect
could also modify the entire inertial range, but this was not confirmed
in subsequent simulations \citep{HB06}.
Furthermore, \cite{DS10} have presented evidence that the bottleneck
becomes less strong at higher Reynolds numbers.
It is therefore perhaps fair to say that the physical reality of the
bottleneck effect is not universally accepted, because it is not easily
seen in high Reynolds number wind tunnel turbulence and in atmospheric
turbulence \citep{Tsuji}.
This is because in wind tunnel turbulence, one only measures a
one-dimensional spectrum.
For pure power laws, one- and three-dimensional power spectra
are identical, but this is not the case near the breakpoint of the
dissipative subrange, where the $k^{-5/3}$ power law changes sharply
into an exponential fall-off.
This causes the bottleneck effect to be greatly diminished in a
one-dimensional projection \citep{DHYB03}.
Thus, we argue that the bottleneck effect is physically real
for the Reynolds numbers under consideration, but that
it is less strong than what is caused by hyperviscosity and it is usually
barely noticeable in one-dimensional measurements.
Furthermore, even though \cite{DS10} find the bottleneck becoming weaker
at larger Reynolds number, this happens slowly and even for a Reynolds
number of 1000 based on the Taylor microscale, the bottleneck is still
rather prominent.

The astrophysical significance of the spectral bottleneck effect lies
in the effect it might potentially have on the numerical study of
small-scale dynamos, at small and moderate $\Pm <1$.
Since these are dynamos that operate on the resistive length scale,
they depend on the small-scale properties of the flow.
Although these scales are in reality very small, this is usually not
the case in most of the numerical simulations that much of our
intuition has to rely upon; see \cite{Cat99} and \cite{TS15} for a
discussion of small-scale dynamos operating near the solar surface.

Specifically, in this paper, we want to address the possibility that the
kinematic small-scale turbulent dynamo at low magnetic Prandtl number
might depend on the forcing scale.
We consider Run~g01 of SB14 which had $\Pm=0.1$ and a magnetic
Reynolds number of 200, so the fluid Reynolds number was 2000.
They used $512^3$ meshpoints, so our first question is whether this
resolution was adequate.
We begin by rerunning their case at a higher resolution of $1152^3$
meshpoints.
We then consider smaller values of the forcing wavenumber, keeping
however the values of $\nu$ and $\eta$ unchanged.

\section{The model}
\label{Model}

Similar to SB14, we consider dynamo action in a cubic domain of size
$L_1^3$, driven by turbulence forced at normalized wavenumbers $\kf/k_1$
ranging from 1.5 to 4, where $k_1=2\pi/L_1$ is the smallest wavenumber
in the domain.
However, unlike SB14, we consider only nonhelical forcing,
so no large-scale dynamo of $\alpha^2$ type is possible \citep{Mof78}.
We are only interested in the early exponential growth or
decay phase and therefore omit the Lorentz force in the
momentum equation.
We consider an isothermal compressible gas and thus
solve the following hydromagnetic evolution equations
for the magnetic vector potential $\AAA$, the velocity $\uu$,
and the density $\rho$,
\begin{eqnarray}
&& \frac{\partial}{\partial t} \AAA = \uu\times\BB -\eta\mu_{0}\JJ,
\label{eq: induction} \\
&& \frac{\DD}{\DD t} \uu = -\cs^{2}\nab \ln{\rho} + \ff
+ \rho^{-1}\nab\cdot2\nu\rho\SSSS,
\label{eq: momentum} \\
&& \frac{\DD}{\DD t} \ln{\rho} = -\nab\cdot\uu, \label{eq: continuity}
\end{eqnarray}
where $\BB=\nab\times\AAA$ is the magnetic field,
$\JJ=\nab\times\BB/\mu_0$ is the current density,
$\mu_0$ is the vacuum permeability,
$\cs=\const$ is the isothermal sound speed, 
$\DD/\DD t=\partial/\partial t+\uu\cdot\nab$ is the advective time derivative,
$\SSSS$ is the traceless rate-of-strain tensor with components
${\sf S}_{ij}=\frac{1}{2}(u_{i,j}+u_{j,i})-\frac{1}{3}\delta_{ij}\nab\cdot\uu$,
and commas denote partial derivatives.
Energy supply is provided by the forcing function $\ff = \ff(\xx,t)$,
which is random in time and defined as
\EQ
\ff(\xx,t)={\rm Re}\{N\ff_{\kk(t)}\exp[\ii\kk(t)\cdot\xx+\ii\phi(t)]\},
\label{ForcingFunction}
\EN
where $\xx$ is the position vector.
The wavevector $\kk(t)$ and the random phase
$-\pi<\phi(t)\le\pi$ change at every time step, so $\ff(\xx,t)$ is
$\delta$-correlated in time.
Therefore, the normalization factor $N$ has to be proportional to $\delta t^{-1/2}$,
where $\delta t$ is the length of the time step.
On dimensional grounds it is chosen to be
$N=f_0 c_{\rm s}(|\kk|c_{\rm s}/\delta t)^{1/2}$, where $f_0$ is a
nondimensional forcing amplitude.
We choose $f_0=0.02$, which, for our range of Reynolds numbers,
results in a maximum Mach number of about 0.5
and an rms velocity of about 0.13, which is almost the same for all the runs.
At each timestep we select randomly one of many possible wavevectors
in a certain range around a given forcing wavenumber with
average value $k_{\rm f}$.

Our model is governed by several nondimensional parameters.
In addition to the scale separation ratio $\kf/k_1$, introduced above,
there are the magnetic Reynolds and Prandtl numbers
\EQ
\Rm=\urms/\eta\kf,\quad
\Pm=\nu/\eta.
\label{Rey_def}
\EN
These two numbers also define the fluid Reynolds number,
$\Rey=\urms/\nu\kf=\Rm/\Pm$.
The maximum values that can be attained are limited by the numerical
resolution and become more restrictive at larger scale separation.
The calculations have been performed using the {\sc Pencil Code}\footnote{
\url{https://github.com/pencil-code}} at a resolution of
$1152^3$ mesh points, except for Run~D3 where $512^3$ mesh points were used.

\begin{table}\caption{
Parameters of Runs~A--D2 at $1152^3$ meshpoints and
comparison with Run~g01 of SB14 and D3 at $512^3$ meshpoints.
Here, $\tilde{k}_{\rm f}=\kf/k_1$,
$\tilde{\epsilon}_{-4}^{\rm K}=\epsK/(10^{-4}\cs^3k_1)$,
$\tilde{u}_{\rm rms}=\urms/\cs$,
$\tilde{\lambda}_{-3}=\lambda/(10^{-3}\cs k_1)$, and
$\tilde{k}_\nu=k_\nu/k_1$.
}\vspace{12pt}\centerline{\begin{tabular}{cccccrccc}
Run & $\tilde{k}_{\rm f}$ &  $\tilde{\epsilon}_{-4}^{\rm K}$ &
$\tilde{u}_{\rm rms}$ & $\Rm$ & $\tilde{\lambda}_{-3}$ &
$\tilde{k}_\nu$ & $a_{\rm K}$ & $a_\eta$ \\
\hline
g01&4.06  & 3.38 & 0.127 & 196 & 2.9& --- & ---  & ---  \\
A & 4.06  & 3.39 & 0.128 & 197 & 2.5& 536 & 1.00 & 1.05 \\
B & 3.13  & 2.85 & 0.129 & 258 & 3.1& 514 & 1.00 & 1.02 \\
C & 2.23  & 1.91 & 0.128 & 359 & 4.6& 465 & 0.95 & 1.03 \\
D & 1.54  & 1.46 & 0.132 & 536 & 4.2& 435 & 0.95 & 1.00 \\
D2& 1.54  & 1.39 & 0.126 & 178&$-1.9$&194 & ---  & ---  \\ 
D3& 1.54  & 1.33 & 0.131 & 284&$ 1.9$&265 & ---  & ---  \\
\label{Tsummary}\end{tabular}}\end{table}

\section{Simulations}
\label{Simulations}

In the following we present runs for $\Pr_M=0.1$ with
different forcing wavenumbers
$\kf$, where $\kf/k_1$ ranges from 4.06 (the value used in SB14)
down to 1.54 (the averaged value for all the $20$ wavevectors
with lengths between $1$ and $\sqrt{3}$; see \Tab{Tsummary}.)
We determine the growth rate $\lambda$ by plotting the logarithm of the
root-mean-square magnetic field, $\ln\Brms$, versus time
in sound travel times, $t\cs k_1$.
The instantaneous growth rate is $\lambda(t)=\dd\ln\Brms/\dd t$.
To have values of the logarithm close to zero, we normalize by
an arbitrarily chosen reference value, $\Bref$.
Statistical errors have been determined as the largest departure
of any one third of that part of time series where the data is
deemed to be in a steady state.

\begin{figure}\begin{center}
\includegraphics[width=\columnwidth]{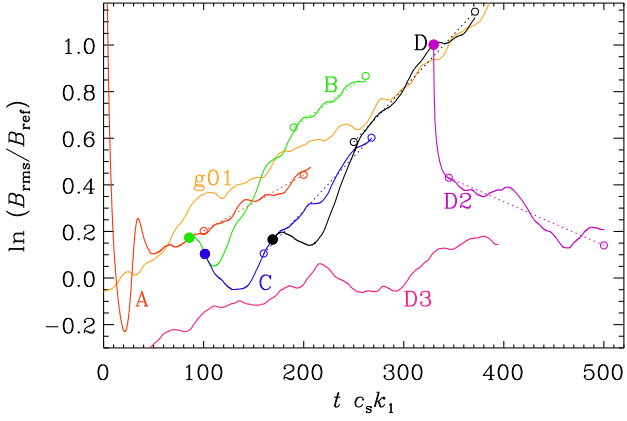}
\end{center}\caption[]{
Time series of $\Brms$ (normalized by an arbitrarily chosen $\Bref$)
for Runs~A--D and comparison with Run~g01 of SB14 (orange).
Except for Run~A, which was started from scratch, Runs~B--D have been
restarted at the times indicated by a filled circle.
Run~D2 (pink) has been restarted with a smaller value of $\Rm$ close to
that of Run~A.
The dotted lines with open circles on their end points indicate the
least-square fit to the last part of the curves indicated.
Run~D3 has been restarted from Run~f01 of SB14 and averaging has been
performed over 800 sound travel times (not indicated).
}\label{pcomp}\end{figure}

It turns out that for $\kf/k_1=4.06$, the growth rate is within
3\% the same as that found by SB14; compare the red line
in \Fig{pcomp} with the orange line for Run~g01.
This is small enough so that we can conclude that their resolution
was adequate.
Nevertheless, we keep this higher resolution for the following study.

We then proceed to lower values of $\kf$.
We expected that at some value the dynamo would cease to be excited.
However, it turned out that, some time after restarting from an earlier
run with larger $\kf$, a clear exponential growth commenced where
the growth rate was even larger than before; see again \Fig{pcomp}.
Note, however, that it always takes some hundred sound travel times
to establish exponential small-scale dynamo growth.
This time corresponds to 7--8 turnover times when based on the values
of $\urms$ and $\kf=1.54\,k_1$ for Run~D, and about 30 turnover times
when $\kf\approx4$ for Run~A.

Of course, given that $\eta$ was unchanged for 
Runs A--D, and $\kf$ and $\urms$ enter in the definition of $\Rm$,
the values of $\Rm$ increase; see \Tab{Tsummary}.
Thus, since Run~D now turns out to be clearly supercritical, we can
conclude that the critical value of $\Rm$ for the small-scale dynamo at
small $\kf$ is clearly below 500.
This is compatible with the simulations of \cite{Isk07}, whose largest
value of $\Rm$ was 450 based on the wavenumber $k_1$ of the domain, and
therefore around 300 when based on $\kf$ (with $\kf=\sqrt{2}\,k_1$
in their case), which is the normalization used in the present paper.

\begin{figure}\begin{center}
\includegraphics[width=\columnwidth]{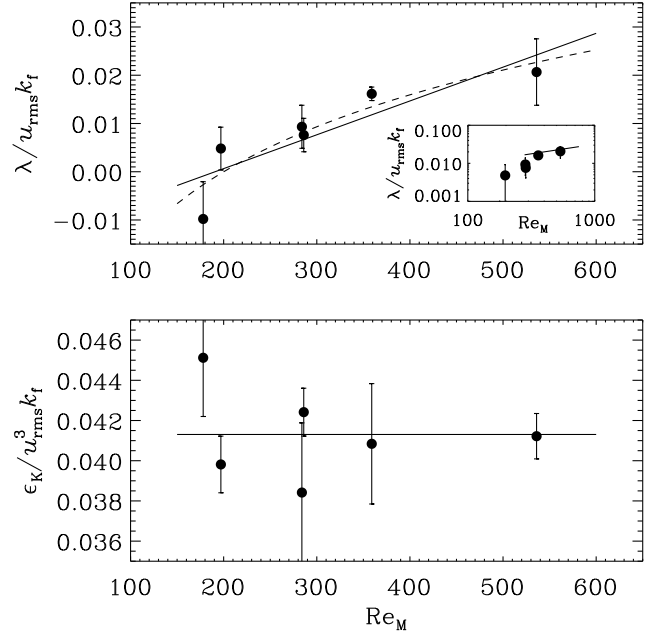}
\end{center}\caption[]{
Growth rate (top) and energy dissipation rate (bottom) versus magnetic
Reynolds number.
In the top panel, the solid and dashed lines refer to \Eqs{linfit}{logfit},
respectively.
The inset shows a comparison with the theoretically
expected $\Rm^{1/2}$ (straight line) scaling in a double-logarithmic
representation.
In the lower panel, the horizontal line gives the average value of 0.040.
}\label{plameps}\end{figure}

Runs~A and D2 have nearly the same $\Rm$, but their growth rates are
not the same.
These two runs do have different values of $\kf/k_1$, so there could be
an additional dependence on the scale separation ratio.
On the other hand, these two data points still fit reasonably well onto
a linear dependence of $\lambda/\urms\kf$ versus $\Rm$; see the top panel
of \Fig{plameps}.
This suggests that the possibility of an additional $\kf$ dependence is
probably not real.
Specifically, we find
\EQ
\lambda/\urms\kf\approx(\Rm-\Rmc)/13,000
\label{linfit}
\EN
with $\Rmc\approx200$.
Furthermore, the largest two data points would be compatible with the
theoretically expected square root dependence,
\EQ
\lambda/\urms\kf\approx10^{-3}\Rm^{1/2}\quad\mbox{(for large $\Rm$)}.
\EN
We also point out that for all these runs, the value of the dissipation
rate, $\epsK=\bra{2\nu\rho\SSSS^2}$, scales well with the theoretical
dependence proportional to $\urms^3\kf$ ($\epsK\approx0.04\,\urms^3\kf$).
Alternatively, for low $\Pm$ dynamos, \cite{KR12} proposed a logarithmic
dependence $\propto\ln(\Rm/\Rmc)$.
Specifically, we find
\EQ
\lambda/\urms\kf\approx0.023\,\ln(\Rm/\Rmc)
\label{logfit}
\EN
as a reasonable fit.

\begin{figure}\begin{center}
\includegraphics[width=\columnwidth]{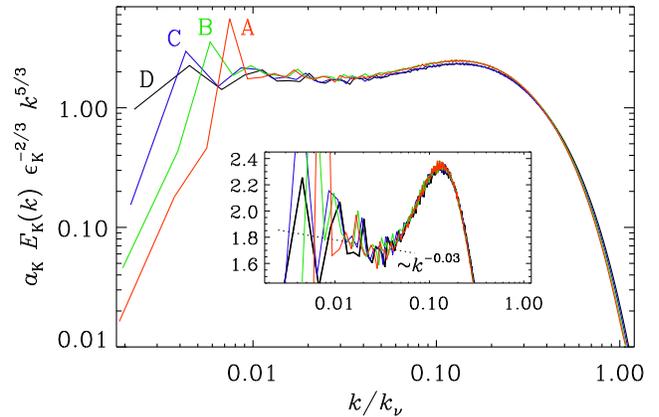}
\end{center}\caption[]{
Compensated kinetic energy spectra for all four runs.
The inset shows the compensated spectra on a linear scale.
The dotted line shows the theoretically expected inertial range
correction proportional to $k^{-0.03}$.
}\label{pspecm_comp}\end{figure}

\begin{figure*}\begin{center}
\includegraphics[width=\textwidth]{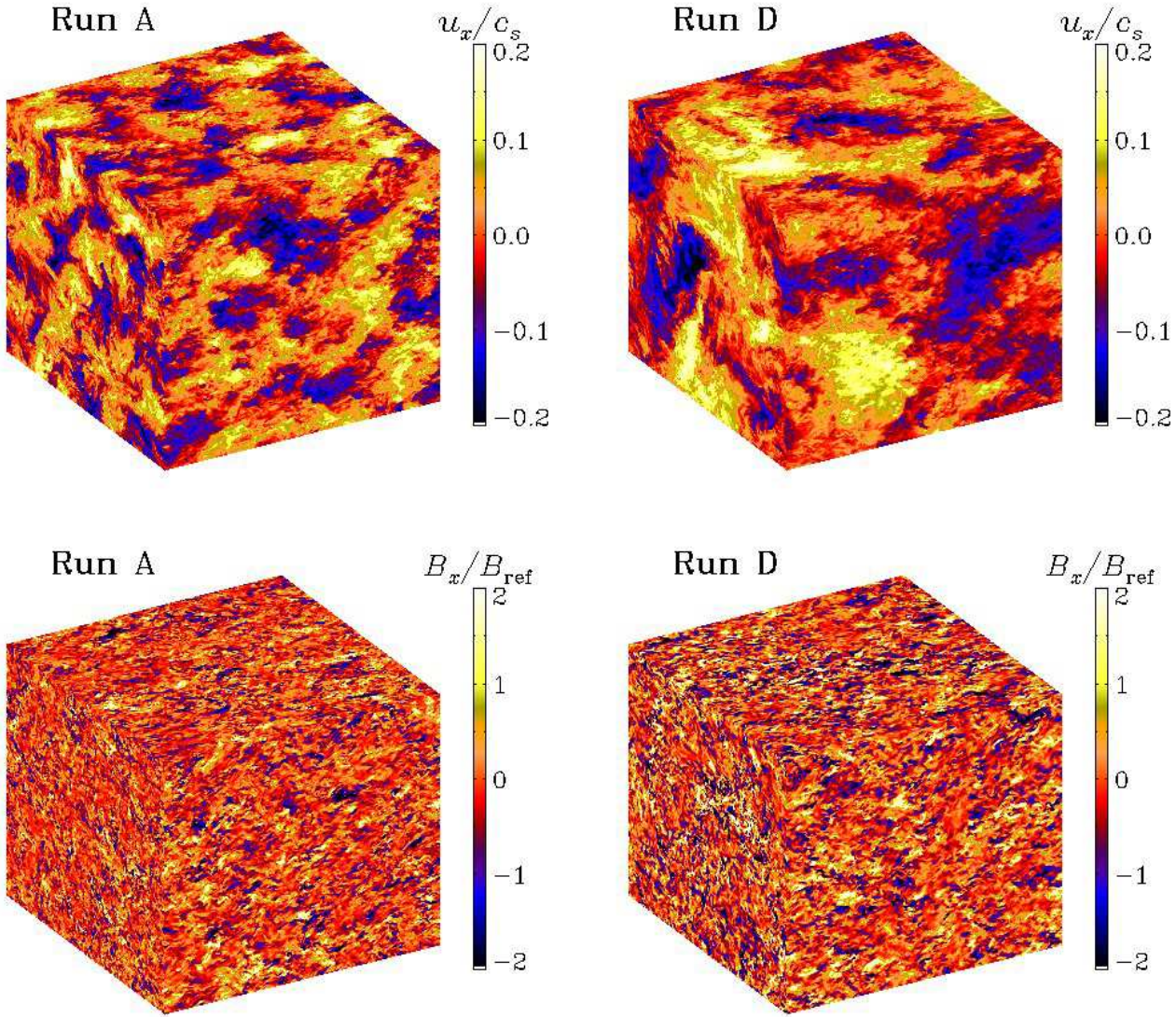}
\end{center}\caption[]{
Visualizations of $u_x$ (top) and $B_x$ (bottom) for Runs~A (left) and D
(right) on the periphery of the domain at the last time of each run.
}\label{UB}\end{figure*}

Let us now discuss the question how to resolve the apparent
conflict between our value of $\Rmc\approx200$ and that
of around 300 obtained by \cite{Isk07}
for a supercritical $\Pm=0.1$ dynamo.
First, they only quote growth for $\Rm=450/\sqrt{2}\approx320$ and decay for
$\Rm=230/\sqrt{2}\approx160$.
Second, interpolating between Runs~D and D2 would yield $\Rmc\approx290$,
which is close to the value of \cite{Isk07}.
Interpolating between Runs~D2 and D3 would yield $\Rmc\approx230$,
but Run~D3 has lower resolution and may be unreliable.
Furthermore, looking again at \Fig{pcomp}, it is clear that there can
sometimes be extended intervals during which the instantaneous growth
rate can be significantly different.
It may however be noted that for $\Pm=1$, the discrepancy is smaller
in that Iskakov et al.\ (2007) found $\Rmc\approx60/\sqrt{2}=42$ while
\cite{HBD04} found $\approx35$.
Third, and this is perhaps the simplest explanation, the effective value
of $\kf$ may not be equal to the nominal average given by $\kf=1.54\,k_1$,
but it may be by up to $300/200$ times larger.
Indeed, if we were to declare that $\kf\to\kf^{\rm eff}\approx2\,k_1$
for Run~D2, it would also resolve the otherwise increasing discrepancy
between Runs~A and D2, which have similar nominal $\Rm$, but different
growth rates.
We therefore propose $\Rmc\approx200$ as the currently best value for
the small-scale dynamo at $\Pm=0.1$.

Next, we study whether the height of the bottleneck was affected as
we changed $\kf$ from 4.06 to 1.54.
The results for the kinetic energy spectra, time-averaged over the
statistically steady state, are shown in \Fig{pspecm_comp}.
These spectra have been compensated by $\epsK^{-2/3}k^{5/3}$.
There we see a short plateau corresponding to the inertial range with
a value (known as the Kolmogorov constant) of around 1.7, close to that
found by \cite{DS10}.
The wavenumber on the abscissa has been scaled by the dissipative
Kolmogorov cutoff wavenumber, $k_\nu=(\epsK/\nu^3)^{1/4}$.
However, in order to achieve ever better overlap between the different
spectra, we have applied an additional scaling factor $a_{\rm K}$
close to unity,
whose values are listed in \Tab{Tsummary} for each of the four runs.

All four curves are seen to collapse perfectly on top of each other at
high wavenumbers.
We can thus conclude that the spectral bottleneck is insensitive to the
details of the large-scale forcing.
This appears plausible, but it was never demonstrated and SB14 were
clearly concerned about this aspect, which is why they considered, for
safety reasons, the value $\kf/k_1\approx4$ instead of just 1.54.
But now we know that this would not have been necessary and that they
could have reached even larger values of $\Rm$ by lowering $\kf$.
However, their findings concerning the small-scale dynamo emerged as a
by-product while studying the kinematic $\alpha^2$ large-scale dynamo in
the presence of helicity.
The fastest growing mode of this dynamo occurs at $k<\kf/2$, which
justified their choice of $\kf/k_1\approx4$ in those cases.

\begin{figure}\begin{center}
\includegraphics[width=\columnwidth]{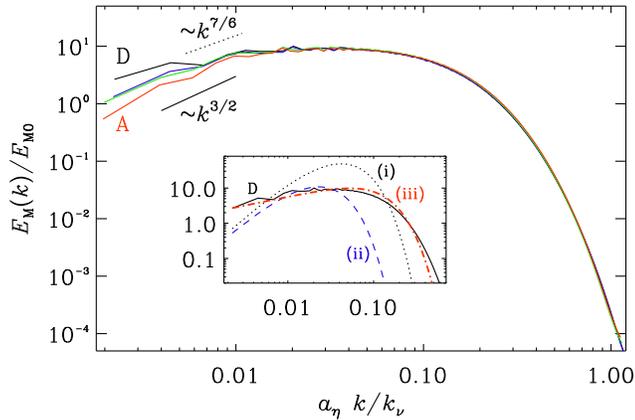}
\end{center}\caption[]{
Magnetic energy spectra for all four runs, time-averaged after
compensating against the exponential growth, as explained in the text.
The coloring of the lines is the same as in \Figs{pcomp}{pspecm_comp}.
The inset shows a comparison between the magnetic spectrum for Run~D
and the Macdonald function with different arguments explained in the text.
}\label{pspecm_kazan}\end{figure}

The inset of \Fig{pspecm_comp} shows more clearly the relevant part
of the inertial range and the bottleneck.
The height of the bottleneck is roughly compatible with what was found
by \cite{Kaneda} using incompressible hydrodynamic turbulence
simulations at a resolution of $4096^3$ meshpoints.
We also see that an inertial range correction of about $k^{-0.03}$,
as expected from the intermittency model of \cite{SL94}, is
compatible with the data.
This is indicated in the inset of \Fig{pspecm_comp} by the dotted line.
Note, however, that the simulations of \cite{Kaneda} and also those
of \cite{HB06} showed a steeper inertial range correction of about
$k^{-0.1}$, which is not theoretically expected.

In \Fig{UB} we compare visualizations of velocity and magnetic
field for Runs~A and D.
For both fields, we show the $x$component on the periphery of the domain.
For both fields, the visual impression is dominated by the largest
structures, especially for the velocity field.
One sees that the smallest structures of the magnetic field tend
to cluster in particular regions in space, especially for Run~D.
Nevertheless, the scale of the magnetic structures seems the same
in both cases.

To verify the visual impression regarding the similarity in scales,
we show in \Fig{pspecm_kazan} magnetic power spectra,
$E_{\rm M}(k)$, which have been averaged after compensating against the
exponential growth with a factor of $\exp(-\lambda t)$, where the values
of $\lambda$ are listed in \Tab{Tsummary}.
The total magnetic energy, $E_{\rm M0}=\int E_{\rm M}(k)\,{\rm d}k$
is used for normalization.
The results shown in \Fig{pspecm_kazan} demonstrate again perfect agreement
between all the spectra at sufficiently large values of $k$.
Again, to achieve better overlap between the different spectra, we have
applied an additional scaling factor $a_\eta$ on the abscissa.
Their values are listed in \Tab{Tsummary}.
Note, however, that the theoretically expected $k^{3/2}$ scaling
seen in simulations with larger values of $\Pm$ \citep{HBD04}
applies at best only to Run~A, and this only over a rather short range.
A shallower $k^{7/6}$ scaling was proposed by SB14 and \cite{BSB16},
but even that appears still too steep for Run~D; see \Fig{pspecm_kazan}.

For $\Pm\gg1$, the theoretically expected eigenfunction is proportional
to $k^{3/2} K_{\nu}(k/k_\eta)$ with index $\nu=0$ and
$k_\eta/k_1=(R_{\rm m}/6)^{1/2}\approx9$ \citep{KA92,Scheko02}.
This was found to provide a good fit to the numerical simulation
of \cite{BS05}.
For low $\Pm$, the functional form yields a much more extended $k^{3/2}$
range than what is seen in the present simulation; see (i) in the inset
of \Fig{pspecm_kazan}.
Lowering it to $k_\eta/k_1=5$ fits the low $k$ range better, but provides
a poor description for large $k$ (ii).
A better overall description is given by $\nu=2$ and
$k_\eta/k_1=(R_{\rm m}/1.4)^{1/2}\approx20$; see (iii) in
the inset of \Fig{pspecm_kazan}.

Finally, let us ask if the above results on the low $\Pm$ dynamo 
can be understood based on general arguments? 
We recall that in the bottleneck region, the spectrum
becomes shallower and the velocity becomes even more rough than
for a pure Kolmogorov spectrum.
Suppose this occurs at a wavenumber around $k_B = f k_\nu$.
At low $\Pm$, the turbulent flow with the resistive scale $k_\eta$ around $k_B$,
will be expected to fail to be a dynamo.
In fact, from the \citet{Kaz68} analysis one knows that a rough velocity with turbulent diffusion
$v_l l \propto l^{1+\alpha}$ and $\alpha < 0$, fails to be a dynamo
\citep{BC04}. This would happen if the resistive scale falls
in the bottleneck region.
For a Kolmogorov spectrum, we have $k_\eta \sim \kf \Rm^{3/4}$, while
$k_\nu \sim \kf \Rey^{3/4}$, and thus 
we expect the low $\Pm$ dynamo to be difficult to excite when
$k_\eta = \kf \Rm^{3/4} = k_B = f k_\nu = f \kf \Rey^{3/4}$, or
when $\Pm = \Rm/\Rey \sim  f^{4/3}$, {\it independent} of $\kf$.
From \Fig{pspecm_comp} we see that the rising part of the bottleneck 
region, where one expects $\alpha\sim 0$, occurs at a wavenumber 
$\sim 0.1\,k_\nu$,
which gives $f \sim 0.1$, implying for the critical $\Pm \sim 0.05$.
This is indeed roughly the value of $\Pm$ where it has been difficult to
excite a small-scale dynamo. The above argument also shows that such
a difficulty should not arise for either a much smaller or larger $\Pm$. 

\section{Conclusions}

Our work has conclusively demonstrated that the inertial range and the
spectral bottleneck of turbulence are not affected by the details of the
forcing at large length scales, specifically the value of $\kf$.
We have also shown that the value of $\kf$ affects neither
the excitation condition of the small $\Pm$ dynamo nor the
shape of the exponentially growing magnetic energy spectrum at
moderate to large values of the wavenumber.
This is significant for future studies of small-scale dynamos in that
it can now be regarded as safe to use the smallest possible forcing
wavenumber in order to maximize the extent of the inertial range.
Clearly, this conclusion does not carry over to other types of dynamos,
notably the large-scale dynamos, where a minimal forcing wavenumber
of about three was found to be just acceptable \citep{BRRS08}.

We have changed $\kf$ only by a factor of less than three, so
we cannot make any claims about larger ranges.
However, when changing $\kf$ from $4.06\,k_1$ (where $210$ different $\kk$
vectors contribute in the range $3.5<|\kk|/k_1<4.5$) to the smallest
possible value of $1.54\,k_1$ (where only $20$ vectors contribute in
$1.4<|\kk|/k_1<1.8$), we have not seen any systematic changes in the
velocity spectrum.
The changes in the growth rate appear to be well explained by the
expected variation of $\Rm$.
For $\kf=1.54\,k_1$, however, we have argued that the effective forcing
wavenumber is $\kf^{\rm eff}\approx2\,k_1$.
It would be of interest to extend our work to values of $\Pm$ well below
$0.1$, to see whether one can find an asymptotic regime where $\Rmc$
becomes independent of $\Pm$.
At the present time, this would be an expensive task, but the day will
come when this can easily be done.

We have seen that for $\Pm=0.1$, the shape of the magnetic energy
spectrum is clearly different from that at large $\Pm$.
It is shallower at large wavenumbers and not proportional to $k^{3/2}$,
as already noted earlier \citep{BSB16}.
Although \cite{KR12} have obtained the real space eigenfunction 
in various asymptotic domains for this case in
analytic form, it still remains to be shown how the spectrum actually
looks like.

Further work on small-scale dynamos remains manifold.
For example, compressibility effects in general and the Mach number
dependence are important \citep{HBM04,Fed11,Fed14,SBS18} and would be interesting
to reconsider at high resolution using direct numerical simulations.
There are also questions regarding the importance of small-scale
dynamo-generated magnetic fields in enhancing the effective turbulent
viscosity, for example in the Sun's convection zone.
Small-scale dynamo-produced magnetic fields may also modify the values
of turbulent transport coefficients, such as turbulent diffusivity and
the $\alpha$ effect \citep{RB10}.
Finally, small-scale dynamo action is known to affect the negative
effective magnetic pressure instability in hydromagnetic turbulence in
the presence of strong density stratification; see \cite{BRK16} for a
recent review of its theory and applications.
Again, more systematic work in that direction is required and would
benefit from the assurance that the dynamo effect is itself not being
affected by the forcing details on the scale of the simulation domain.
However, looking for secondary instabilities, such as the 
negative effective magnetic pressure instability or the large-scale dynamo
instability, obviously requires continued care.

\section*{Acknowledgements}
We thank the referee for inspiring comments that have led to
improvements in the paper.
This work was supported through the FRINATEK grant 231444 under the
Research Council of Norway, the Swedish e-Science Research Centre,
the National Science Foundation, grant AAG-1615100,
the University of Colorado through its support of the
George Ellery Hale visiting faculty appointment,
and the grant ``Bottlenecks for particle growth in turbulent aerosols''
from the Knut and Alice Wallenberg Foundation, Dnr.\ KAW 2014.0048.
The simulations were performed using resources provided by
the Swedish National Infrastructure for Computing (SNIC)
at the Royal Institute of Technology in Stockholm and
Chalmers Centre for Computational Science and Engineering (C3SE).


\vfill\bigskip\noindent\tiny\begin{verbatim}
$Header: /var/cvs/brandenb/tex/mhd/lowPm/paper.tex,v 1.87 2018/06/06 10:36:22 brandenb Exp $
\end{verbatim}

\end{document}